\documentstyle[aps,prl,cite,epsfig,enumerate,verbatim]{revtex}
\draft
\begin{document}
\draft
\title{Degree of entanglement for two qubits}
 \author{
 Jing-Ling Chen,$^{1,2}$ \footnote{Email address: jinglingchen@eyou.com}
 Libin Fu,$^{1}$ Abraham A. Ungar,$^3$
 \footnote{Email address: ungar@gyro.math.ndsu.nodak.edu}
 and Xian-Geng Zhao$^1$}
\address{
 $^1$
 Laboratory of Computational Physics,\\
 Institute of Applied Physics and Computational Mathematics, \\
 P.O. Box 8009(26), Beijing 100088, People's Republic of China. \\
 $^2$
 Department of Physics, Faculty of Science, National University of
 Singapore, \\ Lower Kent Ridge, Singapore 119260, Republic of
 Singapore. \\
 $^3$ Department of Mathematics, North Dakota State University,
 Fargo, North Dakota 58105, USA.}
 \maketitle

 \begin{abstract}

 In this paper, we present a measure to quantify the
 degree of entanglement for two qubits in a pure state.

 \end{abstract}


 \section{Introduction}

 Quantum entanglement is the most surprising nonclassical property
 of composite quantum systems \cite{Schro}. As it is well-known,
 a qubit (or a spin-$1/2$ particle) is described by the $2\times 2$
 density matrix $ \rho({\bf n}) = ({\bf 1}+ {\vec \sigma}
 \cdot {\bf n})/2$, $|{\bf n}| \le 1,$
 where ${\bf 1}$ is the unit matrix,
 ${\vec \sigma}=(\sigma_1, \sigma_2, \sigma_3)$ the Pauli
 matrices vector, and ${\bf n}$ the Bloch vector. $|{\bf n}|=1$
 corresponds to a pure state, otherwise a mixed state. Whereas,
 an entangled pairs of two qubits is completely described by the
 following $4 \times 4$ density matrix:

  \begin{equation}
   \rho_{AB}^{\phantom{2}} = \frac{1}{4} ( {\bf 1} \otimes {\bf 1} +
   {\vec \sigma}^A \cdot {\bf u} \otimes {\bf 1}+
   {\bf 1}\otimes {\vec \sigma}^B \cdot {\bf v}+
   \sum_{i,j=1}^3 \beta_{ij} \sigma_i^A \otimes \sigma_j^B ),
    \label{e1}
 \end{equation}
from which one could obtain two reduced density matrices

 \begin{eqnarray}
&&   \rho_A^{\phantom{2}} = {\rm tr}_B (\rho_{AB}^{\phantom{2}})=
\frac{1}{2} ({\bf 1}+ {\vec \sigma}^A \cdot {\bf u}),\nonumber\\
&&   \rho_B^{\phantom{2}} = {\rm tr}_A (\rho_{AB}^{\phantom{2}})
= \frac{1}{2} ({\bf 1}+ {\vec \sigma}^B \cdot {\bf v}),
 \label{e2}
 \end{eqnarray}
 for the two qubits A and B, where ${\bf u}$ and ${\bf v}$ are
 Bloch vectors for particles A and B, respectively; $\beta_{ij}$
 are some real numbers.

 It has been shown that entangled pairs are a more powerful resource
 than separable, i.e., disentangled, pairs in a number of applications,
 such as quantum cryptography \cite{Ekert}, dense coding \cite{Benne1},
 teleportation \cite{Benne2} and investigations of quantum channels
 \cite{Fisch}, communication protocols and computation \cite{Lo}
 \cite{Niels}. The superior potentiality of
 entangled states has raised a natural question: `` How much are
 two particles entangled?", since pairs with a high degree of
 entanglement should be a better resource than less entangled
 ones. Many measures of entanglement proposed in the past have
 relied on either the Schmidt decomposition \cite{Peres}
 or decomposition in a {\it magic basis} \cite{Woott}.
 In an interesting paper, Abouraddy {\it et al.}
 devised a new measure of entanglement for pure bipartite states
 of two qubits, based on a decomposition of the state vector as a
 superposition of a maximally entangled state vector and an
 orthogonal factorizable one \cite{Abour}.
 Although there are many such decompositions,
 the weights of the two superposed states are remarkably unique. The
 square of the weight of the maximally entangled state vector
 (i.e., $P_E=p^2$) is then defined as the degree of entanglement for two
 qubits, such a measure is consistent with three measures of entanglement
 previously set forth: maximal violation of Bell's inequality \cite{CHSH},
 concurrence \cite{Woott} and two-particle visibility \cite{Jaeger}.

 The purpose of this paper is to propose a new approach
 to the problem of defining the degree of entanglement for
 two qubits in a pure state. In Sec. II, a new measure is formulated
 to quantify the degree of entanglement. Some examples are given in
 Sec. III. Conclusion and discussion are made in the last section.

 \section{Formalism}

 {\bf Theorem:} If $\rho_{AB}^{\phantom{2}}$ is a pure state, then its degree of
 entanglement $P_E$ is equal to

 \begin{equation}
 P_E = (- {\rm det} \hat{\alpha})^{1/4}
 \label{e3}
 \end{equation}
 where the matrix $\hat{\alpha}$ is

\begin{equation}
\hat{\alpha}= \left(
 \begin{array}{llll}
 1 & v_1 & v_2 & v_3 \\
 u_1 & \beta_{11}  &  \beta_{12} &  \beta_{13} \\
 u_2 &  \beta_{21} &   \beta_{22} &  \beta_{23} \\
 u_3 &   \beta_{31} &   \beta_{32} &  \beta_{33}
 \end{array}
 \right).
 \label{e4}
 \end{equation}

  {\bf Proof:} $\rho_{AB}^{\phantom{2}}$ is a pure state implies that
  $\rho_{AB}^2=
  \rho_{AB}^{\phantom{2}}$, from which one obtains the following
  constraints among $u_i$, $v_i$ and $\beta_{ij}$ $(i,j=1, 2, 3)$:

 \begin{equation}
 u_i= \beta_{i1} v_1 + \beta_{i2} v_2 +\beta_{i3} v_3,
 \label{e5}
 \end{equation}
 \begin{equation}
 v_i= \beta_{1i} u_1 + \beta_{2i} u_2 +\beta_{3i} u_3,
 \label{e6}
 \end{equation}
 \begin{equation}
 \sum_{i,j} \beta^2_{ij}= 3-|{\bf u}|^2 - |{\bf v}|^2,
 \label{e7}
 \end{equation}
 \begin{equation}
 \beta_{ij}= u_i v_j - (-1)^{i+j} M_{ij},
 \label{e8}
 \end{equation}
 where $M_{ij}$ is the algebraic complement of the matrix element
 $\beta_{ij}$ for the following $\hat{\beta}$ matrix:

 \begin{equation}
 \hat{\beta}= \left(
 \begin{array}{lll}
 \beta_{11}  &  \beta_{12} &  \beta_{13} \\
  \beta_{21} &   \beta_{22} &  \beta_{23} \\
   \beta_{31} &   \beta_{32} &  \beta_{33}
 \end{array}
 \right).
 \label{e4.1}
 \end{equation}

 Eqs. (\ref{e5}) and (\ref{e6}) can be recast as
 $\hat\beta {\bf v}={\bf u}$, $\hat\beta^T {\bf u}={\bf  v}$,
 where $T$ represents transpose and ${\bf u}=(u_1, u_2, u_3)^T$.
 An interesting result, i.e., $|{\bf u}|=|{\bf v}|$,
 will be obtained immediately from Eqs. (\ref{e5}) and (\ref{e6})
 for the pure state $\rho_{AB}^{\phantom{2}}$ \cite{Kummer}.
 From Eq. (\ref{e8}) we have

 \begin{eqnarray}
&& \beta^2_{11}+ \beta^2_{12} + \beta^2_{13}
 = \beta_{11}u_1 v_1 + \beta_{12}u_1 v_2 +\beta_{13}u_1 v_3
 \nonumber\\
&& {\;\;\;\;\;\;\;\;\;\;\;\;\;\;\;\;\;\;\;\;\;\;\;\;}
-[ (-1)^{1+1} \beta_{11} M_{11}+ (-1)^{1+2} \beta_{12} M_{12}
 +(-1)^{1+3} \beta_{13} M_{13}].
 \label{e9}
 \end{eqnarray}
Due to ${\rm det} \hat\beta= \beta_{11} M_{11}- \beta_{12} M_{12}
 + \beta_{13} M_{13}$ and Eq. (\ref{e5}), one obtains

 \begin{eqnarray}
&& \beta^2_{11}+ \beta^2_{12} + \beta^2_{13}
 - u_1^2= - {\rm det} \hat\beta.
 \label{e10}
 \end{eqnarray}
Similarly,
\begin{eqnarray}
&& \beta^2_{21}+ \beta^2_{22} + \beta^2_{23}
 - u_2^2= - {\rm det} \hat\beta, \nonumber\\
&& \beta^2_{31}+ \beta^2_{32} + \beta^2_{33}
 - u_3^2= - {\rm det} \hat\beta.
 \label{e11}
 \end{eqnarray}
 After combining Eqs. (\ref{e7}), (\ref{e10}), (\ref{e11}), and taking
 $|{\bf u}|=|{\bf v}|$ into account, one easily obtains
 $- {\rm det} \hat\beta= 1 -|{\bf u}|^2$ \cite{Kummer}.
 Consequently, we have

 \begin{eqnarray}
&& (- {\rm det} \hat{\alpha})^{1/4} =
[ (- {\rm det} \hat\beta) (1 -|{\bf u}|^2)]^{1/4}
= \sqrt{1 -|{\bf u}|^2}.
 \label{e12}
 \end{eqnarray}

 One can know from Ref. \cite{Abour} that $P_E= 2 \kappa_1
 \kappa_2$, where $\kappa_1$ and $\kappa_2$ are the two coefficients in
 the Schmidt decomposition $|\Psi \rangle= \kappa_1 |x_1, y_1 \rangle
 + \kappa_2 |x_2, y_2 \rangle$, $\rho_{AB}^{\phantom{2}}
 = |\Psi \rangle \langle
 \Psi|$, where $ \{ |x_1\rangle, |x_2 \rangle \}$ and
 $\{|y_1\rangle, |y_2 \rangle \}$ are orthogonal bases for the
 Hilbert spaces of particles $A$ and $B$, respectively. It is easy to
 prove that
 $\kappa_1= \sqrt{ (1+ |{\bf u}|)/2}$, $\kappa_2=\sqrt{ (1-|{\bf
 u}|)/2}$, which are square-roots of the two eigenvalues of the reduced
 matrix $\rho_A^{\phantom{2}}$ or $\rho_B^{\phantom{2}}$. Therefore we have
 $P_E = (- {\rm det} \hat{\alpha})^{1/4}$. This ends the proof.

 \section{Examples}

 {\it Example 1}. For the state $|\Psi \rangle = (|00 \rangle +
 |01\rangle + |11\rangle )/\sqrt{3}$, one obtains the density
 matrix

 \begin{eqnarray*}
&& \rho_{AB}^{\phantom{2}}= |\Psi \rangle \langle \Psi|= \frac{1}{3}
  \left(
 \begin{array}{llll}
 1  &  1  &  0  &  1  \\
 1  &  1  &  0  &  1  \\
 0  &  0  &  0  &  0  \\
 1  &  1  &  0  &  1
 \end{array}
 \right)
 \label{t1}
 \end{eqnarray*}
 with the Bloch vectors ${\bf u}=( 2/3, 0, 1/3)^T$,
 ${\bf v}=( 2/3, 0, -1/3)^T$, and the alpha matrix

 \begin{equation}
\hat{\alpha}= \frac{1}{3} \left(
 \begin{array}{cccc}
  3& 2 & 0 & -1 \\
  2& 2 &  0  &  -2 \\
  0& 0 &  -2 &  0  \\
  1& 2 &  0  &  1
 \end{array}
 \right).
 \label{t2}
 \end{equation}
 One can have $P_E = 2/3$, which is consistent with the result
 in Ref. \cite{Abour}.

{\it Example 2}. For the state $|\Psi \rangle = [|00 \rangle +
 2 (|01\rangle +  |11\rangle )]/3$, the density matrix is

 \begin{eqnarray*}
&& \rho_{AB}^{\phantom{2}}= |\Psi \rangle \langle \Psi|= \frac{1}{9}
  \left(
 \begin{array}{llll}
 1  &  2  &  0  &  2  \\
 2  &  4  &  0  &  4  \\
 0  &  0  &  0  &  0  \\
 2  &  4  &  0  &  4
 \end{array}
 \right)
 \end{eqnarray*}
 with ${\bf u}=( 8/9, 0, 1/9)^T$,
 ${\bf v}=( 4/9, 0, -7/9)^T$, and the alpha matrix

 \begin{equation}
\hat{\alpha}= \frac{1}{9} \left(
 \begin{array}{cccc}
9 & 4 & 0 & -7 \\
8 & 4 &  0  &  -8 \\
0 & 0 &  -4 &  0  \\
1 & 4 &  0  &  1
 \end{array}
 \right).
 \end{equation}
 Hence the degree of entanglement is $P_E = 4/9$.

 {\it Example 3}. For the maximally entangled state
  $|\Psi \rangle = (|00 \rangle + |11\rangle )/\sqrt{2}$, one obtains
  the density matrix

 \begin{eqnarray*}
 \rho_{AB}^{\phantom{2}}= |\Psi \rangle \langle \Psi|= \frac{1}{2}
  \left(
 \begin{array}{llll}
 1  &  0  &  0  &  1  \\
 0  &  0  &  0  &  0  \\
 0  &  0  &  0  &  0  \\
 1  &  0  &  0  &  1
 \end{array}
 \right) \nonumber\\
 \label{t3}
 \end{eqnarray*}
 with the Bloch vectors ${\bf u}={\bf v}=( 0, 0, 0)^T$, and the alpha
 matrix

 \begin{equation}
\hat{\alpha}= \left(
 \begin{array}{cccc}
1 & 0& 0 & 0 \\
0 & 1 &  0  &  0 \\
0 &  0 &  -1 &  0  \\
 0&  0 &  0  &  1
 \end{array}
 \right).
 \label{t4}
 \end{equation}
 Thus $P_E = 1$ reaches the highest value.

 {\it Example 4}. For the disentangled pure state
  $ \rho_{AB}^{\phantom{2}}=\frac{1}{2} ({\bf 1}+
   {\vec \sigma}^A \cdot {\bf u})
  \otimes \frac{1}{2} ({\bf 1}+ {\vec \sigma}^B \cdot {\bf v})$,
  where $|{\bf u}|=|{\bf v}|=1$,  we have the alpha matrix as

 \begin{equation}
\hat{\alpha}= \left(
 \begin{array}{cccc}
 1 & v_1 & v_2 & v_3 \\
 u_1& u_1 v_1 &  u_1 v_2  &  u_1 v_3 \\
 u_2 & u_2 v_1 &  u_2 v_2  &  u_2 v_3 \\
 u_3 & u_3 v_1 &  u_3 v_2  &  u_3 v_3
\end{array}
 \right).
 \label{t5}
 \end{equation}
 Obviously $P_E =0$ indicates that $\rho_{AB}^{\phantom{2}}$ is disentangled.

 \section{Conclusion and Discussion}

     In conclusion, we have presented a measure to
 quantify the degree of entanglement for two qubits in a pure
 state. We would like to make some discussion in the following:

 (i) The similar idea developed in this paper could be
 generalized to quantify the
 degree of entanglement for two qu$N$its (i.e., $N$-state quantum
 systems, $N=2$ and $N=3$ correspond to a qubit and a qutrit,
 respectively) \cite{Dago}\cite{Milburn} in a pure state. For instance,
 the density matrix for two entangled qutrits could be written as

 \begin{equation}
   \rho_{AB}^{\phantom{2}}= \frac{1}{9} ( {\bf 1} \otimes {\bf 1} +
   \sqrt{3} {\vec \lambda}^A \cdot {\bf u} \otimes {\bf 1}+
   \sqrt{3} {\bf 1}\otimes {\vec \lambda}^B \cdot {\bf v}+
   \frac{3}{2} \sum_{i,j=1}^8 \beta_{ij} \lambda_i^A \otimes \lambda_j^B ),
    \label{c1}
 \end{equation}
 where $\lambda_i$ $(i=1,2,...,8)$ are the eight hermitian
 generators of $SU(3)$ (namely the usual Gell-mann matrices). For the
 state of two entangled qutrits
 \begin{equation}
 |\Psi \rangle = \frac{1}{3} (|00 \rangle + |11\rangle +
 |22\rangle),
  \label{c2}
 \end{equation}
 its corresponding density matrix is \cite{Milburn}
 \begin{equation}
   \rho_{AB}^{\phantom{2}}= \frac{1}{9} ( {\bf 1} \otimes {\bf 1} +
   \frac{3}{2} \sum_{i,j=1}^8 \beta_{ij} \lambda_i^A \otimes \lambda_j^B ),
  \label{c3}
 \end{equation}
 with the non-zero coefficients $\beta_{11}=\beta_{33}=\beta_{44}
 =\beta_{66}=\beta_{88}=1$, $\beta_{22}=\beta_{55}=\beta_{77}=-1$.
 The elements $\beta_{ij}$, 1, ${\bf u}$ and ${\bf v}$ form a $9 \times 9$
 matrix $\hat\alpha$,
 it is easy to show that $P_E=(- {\rm det} \hat{\alpha})^{1/4}=1$,
 which indicates that the state $| \Psi \rangle$ in Eq. (\ref{c2})
 is just a maximally entangled state.

 (ii) After making the parametrization
 ${\bf u}=\hat{\bf u} \tanh \phi_{\bf u}$,
 where $\hat{\bf u}={\bf u}/|{\bf u}|$,
 the density matrix of a qubit
 $\rho({\bf u})= ( {\bf 1}+ {\vec \sigma} \cdot {\bf u})/2$ can be
 connected to the Lorentz boost matrix
 $L({\bf u})=\exp(\varphi_{\bf u} \vec{\sigma}\cdot\hat{\bf u}/2)
 = {\bf 1} \cosh(\varphi_{\bf u}/2)+ {\vec
 \sigma}\cdot{\hat{\bf u}}\sinh(\varphi_{\bf u}/2)$ as \cite{Chenjl1}
 \begin{equation}
 \rho({\bf u})= \frac{L({\bf u})}{ 2 \cosh\phi_{\bf u}},\;\;\;
 \phi_{\bf u}  =\varphi_{\bf u}/2.
 \end{equation}
 Obviously, $\rho({\bf u})$ and $L({\bf u})$ are in one-to-one correspondence.
 For the former, the physical meaning of the vector ${\bf u}$ is the Bloch
 vector in quantum mechanics, while for the latter the relativistic velocity.
 Due to the rapidity $\varphi$, i.e., the hyperbolic angle, special
 relativity can be formulated in terms of hyperbolic geometry. As a result,
 some physical quantities have been found to have geometric
 significance, such as the Thomas rotation angle corresponds to the
 defect of a hyperbolic triangle \cite{Chenjl2}\cite{Ungar}.
 After viewing the Bloch vector ${\bf u}$ as an analogous relativistic
 velocity, the Bures fidelity $F(\rho_1,\rho_2)=
 [{\rm tr} \sqrt{ \sqrt{\rho_1} \rho_2 \sqrt{\rho_1}} ]^2$
 was found to have a geometric interpretation
 in the framework of hyperbolic geometry \cite{Chenjl1}.
 Similarly, with the aid of the parametrization
 ${\bf u}=\hat{\bf u} \tanh \phi_{\bf u}$, it is not difficult to
 find that the entanglement degree
 $P_E=\sqrt{1-|{\bf u}|^2}=1/\cosh\phi_{\bf u}$
 for two qubits in a pure state is the reciprocal of the
 Lorentz factor \cite{Ungar} in the hyperbolic geometry.
 The extension of our approach to the mixed states of two
 entangled qubits will be discussed elsewhere.


\begin{thebibliography}{99}

\bibitem{Schro}  E. Schr{\" o}dinger, Naturwissenschafen {\bf 23},
             807 (1935); {\bf 23}, 823 (1935); {\bf 23}, 844 (1935);
             A. Einstein, B. Podolsky and N. Rosen, Phys. Rev. {\bf 47},
             777 (1935); J.S. Bell, Physics (Long Island, NY),
             {\bf 1}, 195 (164); reprinted in J.S. Bell, {\it
             Speakable and Unspeakable in Quantum Mechanics},
             (Cambridge University Press, Cambridge, 1987).
\bibitem{Ekert}  A.K. Ekert, Phys. Rev. Lett. {\bf 67}, 661 (1991).
\bibitem{Benne1} C.H. Bennett and S.J. Wiesner, Phys. Rev. Lett. {\bf 69},
             2881 (1992).
\bibitem{Benne2} The theoretical proposals are in C.H. Bennnett
                 {\it et al.}, Phys. Rev. Lett. {\bf 70}, 1895
                 (1993); L. Davidovich {\it et al.}, Phys. Rev. A
                 {\bf 50}, R895 (1994). For the experiments see
                 D. Boschi {\it et al.}, Phys. Rev. Lett. {\bf 80}, 1121
                 (1998); A. Furusawa {\it et al.}, Science {\bf
                 282}, 706 (1998); M.A. Nielsen, E. Knill and R.
                 Laflamme, Nature {\bf 396}, 52 (1998).
\bibitem{Fisch}  D.G. Fischer, H. Mack, M.A. Cirone and M. Freyberger,
                 Phys. Rev. A {\bf 64}, 022309 (2001); C.
                 Machiavello and G.M. Palma, e-print quant-ph/0107052.
\bibitem{Lo} H.-K. Lo, T. Spiller and S. Popescu (eds.), {\it Introduction
             to Quantum Computation and Information} (World Scientific
             Publishing, Singapore, 1998); J. Gruska, {\it Quantum
             Computing} (McGraw-Hill, London, 1999); G. Alber {\it
             et al.}, {\it Quantum Information: An Introduction to
             Basic Theoretical Concepts and Experiments}
             (Springer, Berlin, 2001).
\bibitem{Niels} M.A. Nielsen and I.L. Chuang, {\it Quantum
             Computation and Quantum Information} (Cambridge University
             Press, 2000); D. Bouwmeester, A. Ekert and A.
             Zeilinger (eds.), {\it The Physics of Quantum
             Information} (Springer, Berlin, 2000).
\bibitem{Peres} A. Peres, {\it Quantum Theory: Concepts and
             Methods} (Kluwer, The Netherlands, 1993);
             M.A. Cirone, e-print quant-ph/0110139.
\bibitem{Woott} S. Hill and W.K. Wootters, Phys. Rev. Lett. {\bf
             78}, 5022 (1997); W.K. Wootters, Phys. Rev. Lett.
             {\bf 80}, 2245 (1998); K.M. O'Connor and W.K. Wootters,
             Phys. Rev. A {\bf 63}, 052302 (2001).
\bibitem{Abour} A.F. Abouraddy, B.E.A. Saleh, A.V. Sergienko and
             M.C. Teich, Phys. Rev. A {\bf 64}, 050101(R) (2001);
             e-print quant-ph/0109081.
\bibitem{CHSH} J.F. Clauser, M.A. Horne, A. Shimony and R.A. Holt,
             Phys. Rev. Lett. {\bf 23}, 880 (1969);
             N. Gisin, Phys. Lett. A {\bf 154}, 201 (1991);
             S. Popescu and D. Rohrlich, Phys. Lett. A {\bf 166}, 293
             (1992).
\bibitem{Jaeger} G. Jaeger, A. Shimony and L. Vaidman,
             Phys. Rev. A {\bf 51}, 54 (1993).
\bibitem{Kummer} H.J Kummer, Inter. J. Theore. Phys. {\bf 40}, 1071 (2001).
\bibitem{Dago} D. Kaszlikowski, P. Gnaci\'nski, M. \.Zukowski,
             W. Miklaszewski and
             A. Zeilinger, Phys. Rev. Lett. {\bf 85}, 4418 (2000);
             J.-L. Chen, D. Kaszlikowski, L.C. Kwek, M. \.Zukowski,
             and C. H. Oh, Phys. Rev. A {\bf 64}, 052109 (2001).
\bibitem{Milburn} C.M. Caves and G.J. Milburn, e-print quant-ph/9910001;
             P. Rungta, W.J. Munro, K. Nemoto, P. Deuar, G.J.
             Milburn and C.M. Caves, e-print quant-ph/0001075.
\bibitem{Chenjl1}J.-L. Chen, L. Fu, A.A. Ungar and X.-G. Zhao,
             Geometric observation for the Bures fidelity between two
             states of a qubit, Phys. Rev. A, {\bf 65}, 024303 (2001).
\bibitem{Chenjl2}  J.-L. Chen and M.-L. Ge, J. Geom. Phys.
             {\bf 25}, 341 (1998); P.K. Aravind, Am. J. Phys.
             {\bf 65}, 634 (1997);
             A.A. Ungar, Found. Phys. {\bf 27}, 881 (1997);
             J.-L. Chen and A.A. Ungar, Found. Phys.
             {\bf 31}, 1611 (2001).
\bibitem{Ungar}  Abraham A. Ungar, {\it Beyond the Einstein addition law and
             its gyroscopic Thomas precession: the theory of gyrogroups
             and gyrovector spaces} (Kluwer Academic
             Publishers, Dordrecht, 2001).


\end{thebibliography}
 \end{document}